\documentclass[twocolumn,showpacs,preprintnumbers,amsmath,amssymb,aps,prl,superscriptaddress]{revtex4}
\usepackage{color}
\usepackage{graphicx}% Include figure files
\usepackage{dcolumn}% Align table columns on decimal point
\usepackage{bm}% bold math

\begin{document}

\preprint{}

\title{Forward Neutral Pion Transverse Single Spin Asymmetries in p+p Collisions 
at {\boldmath$\sqrt{s}=200\,$}GeV}

\affiliation{Argonne National Laboratory, Argonne, Illinois 60439}
\affiliation{University of Birmingham, Birmingham, United Kingdom}
\affiliation{Brookhaven National Laboratory, Upton, New York 11973}
\affiliation{California Institute of Technology, Pasadena, California 91125}
\affiliation{University of California, Berkeley, California 94720}
\affiliation{University of California, Davis, California 95616}
\affiliation{University of California, Los Angeles, California 90095}
\affiliation{Universidade Estadual de Campinas, Sao Paulo, Brazil}
\affiliation{Carnegie Mellon University, Pittsburgh, Pennsylvania 15213}
\affiliation{University of Illinois at Chicago, Chicago, Illinois 60607}
\affiliation{Creighton University, Omaha, Nebraska 68178}
\affiliation{Nuclear Physics Institute AS CR, 250 68 \v{R}e\v{z}/Prague, Czech Republic}
\affiliation{Laboratory for High Energy (JINR), Dubna, Russia}
\affiliation{Particle Physics Laboratory (JINR), Dubna, Russia}
\affiliation{University of Frankfurt, Frankfurt, Germany}
\affiliation{Institute of Physics, Bhubaneswar 751005, India}
\affiliation{Indian Institute of Technology, Mumbai, India}
\affiliation{Indiana University, Bloomington, Indiana 47408}
\affiliation{Institut de Recherches Subatomiques, Strasbourg, France}
\affiliation{University of Jammu, Jammu 180001, India}
\affiliation{Kent State University, Kent, Ohio 44242}
\affiliation{University of Kentucky, Lexington, Kentucky, 40506-0055}
\affiliation{Institute of Modern Physics, Lanzhou, China}
\affiliation{Lawrence Berkeley National Laboratory, Berkeley, California 94720}
\affiliation{Massachusetts Institute of Technology, Cambridge, MA 02139-4307}
\affiliation{Max-Planck-Institut f\"ur Physik, Munich, Germany}
\affiliation{Michigan State University, East Lansing, Michigan 48824}
\affiliation{Moscow Engineering Physics Institute, Moscow Russia}
\affiliation{City College of New York, New York City, New York 10031}
\affiliation{NIKHEF and Utrecht University, Amsterdam, The Netherlands}
\affiliation{Ohio State University, Columbus, Ohio 43210}
\affiliation{Panjab University, Chandigarh 160014, India}
\affiliation{Pennsylvania State University, University Park, Pennsylvania 16802}
\affiliation{Institute of High Energy Physics, Protvino, Russia}
\affiliation{Purdue University, West Lafayette, Indiana 47907}
\affiliation{Pusan National University, Pusan, Republic of Korea}
\affiliation{University of Rajasthan, Jaipur 302004, India}
\affiliation{Rice University, Houston, Texas 77251}
\affiliation{Universidade de Sao Paulo, Sao Paulo, Brazil}
\affiliation{University of Science \& Technology of China, Hefei 230026, China}
\affiliation{Shanghai Institute of Applied Physics, Shanghai 201800, China}
\affiliation{SUBATECH, Nantes, France}
\affiliation{Texas A\&M University, College Station, Texas 77843}
\affiliation{University of Texas, Austin, Texas 78712}
\affiliation{Tsinghua University, Beijing 100084, China}
\affiliation{Valparaiso University, Valparaiso, Indiana 46383}
\affiliation{Variable Energy Cyclotron Centre, Kolkata 700064, India}
\affiliation{Warsaw University of Technology, Warsaw, Poland}
\affiliation{University of Washington, Seattle, Washington 98195}
\affiliation{Wayne State University, Detroit, Michigan 48201}
\affiliation{Institute of Particle Physics, CCNU (HZNU), Wuhan 430079, China}
\affiliation{Yale University, New Haven, Connecticut 06520}
\affiliation{University of Zagreb, Zagreb, HR-10002, Croatia}

\author{B.I.~Abelev}\affiliation{University of Illinois at Chicago, Chicago, Illinois 60607}
\author{M.M.~Aggarwal}\affiliation{Panjab University, Chandigarh 160014, India}
\author{Z.~Ahammed}\affiliation{Variable Energy Cyclotron Centre, Kolkata 700064, India}
\author{B.D.~Anderson}\affiliation{Kent State University, Kent, Ohio 44242}
\author{D.~Arkhipkin}\affiliation{Particle Physics Laboratory (JINR), Dubna, Russia}
\author{G.S.~Averichev}\affiliation{Laboratory for High Energy (JINR), Dubna, Russia}
\author{Y.~Bai}\affiliation{NIKHEF and Utrecht University, Amsterdam, The Netherlands}
\author{J.~Balewski}\affiliation{Indiana University, Bloomington, Indiana 47408}
\author{O.~Barannikova}\affiliation{University of Illinois at Chicago, Chicago, Illinois 60607}
\author{L.S.~Barnby}\affiliation{University of Birmingham, Birmingham, United Kingdom}
% ...

\collaboration{STAR Collaboration}\homepage{www.star.bnl.gov}\noaffiliation

\date{\today}

\begin{abstract}
We report precision measurements of the Feynman-$x$ ($x_F$) dependence, and
first measurements of the transverse momentum ($p_T$) dependence, of transverse single
spin asymmetries for the production of 
$\pi^0$ mesons from polarized proton collisions at
$\sqrt{s}$=200 GeV.  The $x_F$ dependence of the results are in fair
agreement with perturbative QCD (pQCD) model calculations that identify orbital
motion of quarks and gluons within the proton as the origin of the spin effects. 
Results for the $p_T$ dependence at fixed $x_F$ are 
not consistent with these same pQCD-based calculations.
\end{abstract}

\maketitle

The production of particles with high
transverse momentum from polarized proton collisions at high energies
is sensitive to the quark ($q$) and gluon ($g$) spin structure of the proton.
Perturbative QCD (pQCD) calculations are used to interpret spin
observables when they can explain measured cross
sections.  The goal of measuring spin observables is to understand
how the proton gets its spin from its $q,g$ constituents.

One challenge to theory has been to understand the
azimuthal asymmetry of particles produced in collisions of
transversely polarized protons, known as analyzing power ($A_N$) 
or transverse single spin asymmetry.  With vertical polarization,
non-zero $A_N$ corresponds to a left-right asymmetry of the 
produced particles.  Sizeable $A_N$ are
not expected in collinear pQCD at leading twist due to the chiral
properties of the theory \cite{kane}.  Nonetheless, large $A_N$
are observed for inclusive
pion production in $p_\uparrow+p$ collisions over a broad range of
collision energies ($\sqrt{s}$)
\cite{lowenergy,mediumenergy,largeenergy,E704,starfpd1} and 
in semi-inclusive deep inelastic scattering (SIDIS) from
transversely polarized proton targets \cite{hermes}.  These
observations have prompted extensions to pQCD that
introduce transverse momentum dependence (TMD) correlated with the spin
degree of freedom.  For example, $A_N$ could be generated by spin-correlated TMD fragmentation
if there is transverse $q$ polarization in a transversely
polarized proton (``Collins effect'') \cite{collins1}. 
This mechanism was considered to be suppressed for $p_\uparrow+p\rightarrow\pi + X$
until recently \cite{anselmino1,fyuan}.  Spin-correlated TMD distribution functions (``Sivers functions'')
\cite{sivers,bhs}
can 
explain large $A_N$ \cite{dalesio1}.
These functions describe
parton orbital motion within the proton, and so are
important for understanding the structure of the proton.

Although Sivers functions are
extracted from SIDIS results, there is no proof \cite{collinsqiu} that
they factorize in pQCD calculations of $p_\uparrow+p\rightarrow\pi+X$.
A factorized
framework involving twist-3 $qg$ correlators has been introduced
\cite{qiusterman} and has successfully described \cite{KQVY} previous
$A_N$ results \cite{E704,starfpd1} for $p_\uparrow+p\rightarrow
\pi+X$.  Of relevance to both approaches is a
transverse momentum ($k_T$) that is integrated over in inclusive processes.
This $k_T$ is intrinsic parton motion in the
Sivers functions and its average is related to the inverse proton
radius.  Large $k_T$
is where $qg$ correlators are expected to provide a robust framework.
Small $k_T$ is where Sivers functions are expected to be applicable.
Intermediate $k_T$ values yield the same results in the two
approaches, because moments of the Sivers functions are found to be
related to the $qg$ correlators \cite{BMP,JQVY}.

Both theoretical frameworks \cite{dalesio1,KQVY} predict
that $A_N$ will increase as the longitudinal momentum ($p_L$) of the pion
increases, usually given by Feynman-$x$, $x_F=2p_L/\sqrt{s}$.
Both frameworks predict
that, at fixed $x_F$, $A_N$ will decrease with increasing transverse momentum ($p_T$), for
$p_T>$1.2 GeV/c.

Analyzing powers in the hadroproduction of pions have been measured before,
and typically show a strong increase as
$x_F$ increases
\cite{lowenergy,mediumenergy,largeenergy,E704,starfpd1}.  Virtually no
previous experimental results exist for the dependence of $A_N$ on
$p_T$ at fixed $x_F$. For $\sqrt{s}\le 20$ GeV, the cross sections are
at least ten times larger than pQCD calculations for $x_F$ values
where $A_N$ is sizeable \cite{bs}.  This led to the suggestion that
beam fragmentation, the dissociation of the polarized proton by the
unpolarized target, was responsible for the spin effects, and the
expectation that at sufficiently large $p_T$ these spin effects would
vanish.  At $\sqrt{s}$=200 GeV, inclusive $\pi$ cross sections at
central and forward rapidity are found to be in agreement with pQCD
calculations above $p_T \sim 2$ GeV/c, and are included with world data
for $\pi$ production from $e^+e^-$ collisions, SIDIS, and other $p+p$
collider results in a global analysis of fragmentation functions
\cite{dFSS}.  $A_N$ that increase with $x_F$ are found at
$\sqrt{s}=200$ GeV \cite{starfpd1,bland,brahms}, but both precision measurements
and the determination of the dependence on $p_T$ have, until now, been
missing.

In this Letter, we report precision measurements of the $x_F$ dependence
and first measurements of the $p_T$ dependence of $A_N$ at fixed $x_F$
for $p_\uparrow+p\rightarrow\pi^0+X$ at $\sqrt{s}=200$ GeV. The experiment
has been performed at the Solenoidal Tracker at RHIC (STAR) \cite{star} 
at the Relativistic Heavy Ion
Collider (RHIC) at Brookhaven National Laboratory. The experiment was performed using
vertically polarized colliding beams.  Asymmetries are formed from yields
measured with left-right symmetrical detectors, tagged by the polarization
direction of one beam and summing over the polarization of the other beam.
Positive $x_F$ is probed by considering polarization of the
beam heading towards the detectors and negative $x_F$ is probed by considering
polarization of the beam heading away from the detectors.

Measurements were carried out with a modular electromagnetic calorimeter,
known as the Forward Pion Detector (FPD), positioned at large pseudorapidity
($\eta=-{\rm ln}({\rm tan} \theta/2$)). The $\langle\eta\rangle$=4.0 results, and some
$\langle\eta\rangle$=3.7 results, reported here were obtained in the 2003 (2005) run
having integrated luminosity $L_{int}$=0.25 pb$^{-1}$ (0.1 pb$^{-1}$) and
average beam polarization $P_b\sim$35\% (50\%). $\langle\eta\rangle$=3.3 and
most of the $\langle\eta\rangle$=3.7 measurements were performed in the 2006 run,
which resulted in $L_{int}$=6.8 pb$^{-1}$ with $P_b\sim$55\%.
In the 2006 run, 111 of the 120 possible bunches of both RHIC rings, called
``Blue'' and ``Yellow'', were filled with protons having predetermined patterns
of polarization signs. The unfilled 9 bunches are sequential and correspond
to the abort gap needed to eject the stored beams. 
$P_b$ was measured every 3 hours during RHIC stores by
a polarimeter that detected recoil carbon ions produced in
elastic scattering of protons from carbon ribbon targets inserted into
the beams.  
The effective $A_N$ of this polarimeter was determined from
$p_\uparrow+p_\uparrow$ elastic scattering from a polarized gas jet
target \cite{okada} thereby determining $P_b=55.0\pm 2.6$\% ($56.0\pm 2.6$\%)
for the Blue (Yellow) beam in the 2006 run \cite{bazilevsky}.

The FPD comprises four modules, each containing a matrix of lead glass (PbGl)
cells of dimension 3.8 cm $\times$ 3.8 cm $\times$ 18 radiation lengths. Pairs
of modules were positioned symmetrically left (L) and right (R) of the beamline
in both directions, at a distance of $\sim$750 cm from the interaction point
\cite{bland}. The modules facing the Yellow (Blue) beam are square matrices
of $7 \times 7$ ($6\times 6$) PbGl cells. Data from all FPD cells were
encoded for each bunch crossing, but only recorded when the summed energy
from any module crossed a preset threshold.

\begin{figure}
\includegraphics[height=2.7in]{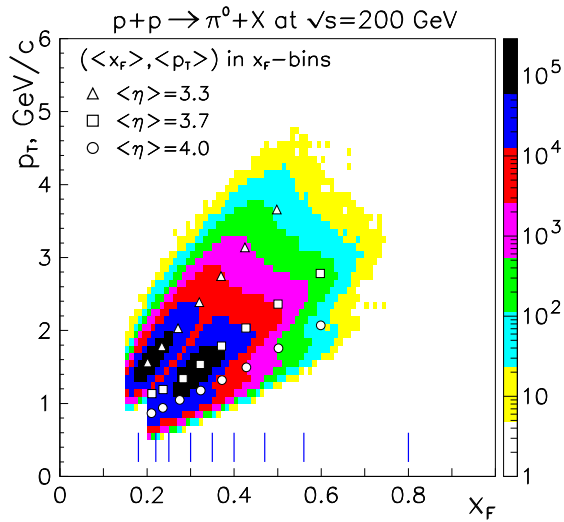}
\caption{Correlation between pion longitudinal momentum scaled
by $\sqrt{s}/2$ ($x_F$) and transverse momentum ($p_T$) for all events.
Bins in $x_F$ used in Figs.~\ref{fig:2} and \ref{fig:4} are indicated by the vertical lines. 
There is a strong correlation between $x_F$ and $p_T$ at a single pseudorapidity
($\langle\eta\rangle$).  
\label{fig:1}}
\end{figure}

\begin{figure}
\includegraphics[height=2.68in]{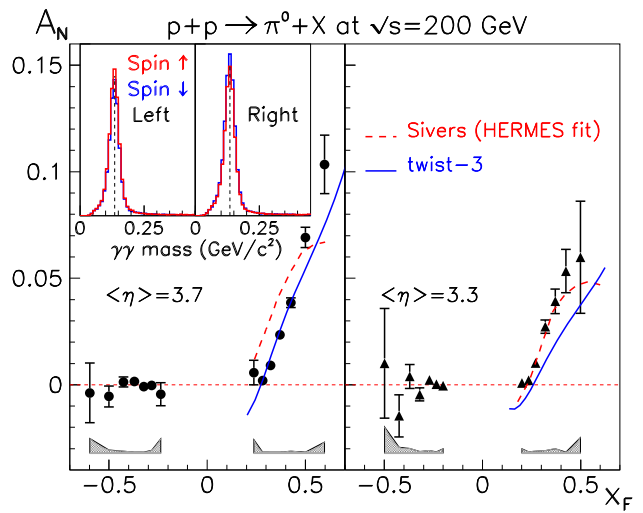}
\caption{Analyzing powers in $x_F$ bins (see Fig.~\ref{fig:1})
at two different $\langle\eta\rangle$.  Statistical errors
are indicated for each point.  Systematic errors are given
by the shaded band, excluding normalization uncertainty.  The 
calculations
are described in the text.  The inset shows examples of
the spin-sorted invariant mass distributions.  The vertical
lines mark the $\pi^0$ mass.
\label{fig:2}}
\end{figure}

Neutral pions are reconstructed via the decay $\pi^0 \rightarrow \gamma\gamma$.
The offline event analysis included conversion of the data to energy
for each cell, formation of clusters and reconstruction of photons using 
a fit with the function that parameterizes the average transverse profile
of electromagnetic showers. Collision events were identified 
by requiring a coincidence between the east and west STAR beam-beam counters (BBC), 
as used for cross section measurements \cite{starfpd2}.
Events were selected when two reconstructed
photons were contained in a fiducial volume, whose boundary excludes a region
of width 1/2 cell at the module edges.
Detector calibration was determined from the $\pi^0$ peak position 
in di-photon invariant mass ($M_{\gamma\gamma}$) distributions. 
The estimated calibration accuracy is 2\%. The analysis was validated 
by checking against full PYTHIA/GEANT simulations \cite{simulation}.
The reconstructed $\pi^0$ energy resolution is given by $\delta E_{\pi}/E_{\pi}\approx
0.16/\sqrt{E_{\pi}}$.

Due to the limited acceptance there is
a strong correlation between $x_F$ and $p_T$ for reconstructed 
$\pi^0$ (Fig.~\ref{fig:1}).  Spin effects in
the $x_F$--$p_T$ plane are studied by positioning the 
calorimeters at different transverse distances from the beam,
maintaining L/R symmetry for pairs of modules.
Fig.~\ref{fig:1} shows loci from $\langle\eta\rangle$=3.3, 3.7 and 4.0.  
There is overlap between the loci, providing cross
checks between the measurements. Because the measurements were made
at a colliding beam facility both $x_F>0$ and $x_F<0$ results are
obtained concurrently. 

Events with 0.08 $<M_{\gamma\gamma}<$ 0.19 GeV/c$^2$ were counted separately by spin state 
from one or the other beam, with no condition on the spin state of the second beam, 
in the $x_F$ bins shown in Fig.~\ref{fig:1}.  
For each run $i$, $A_{N,i}$ for each bin was then
determined by forming a cross ratio
\begin{equation}
A_{N,i} = \frac{1}{P_b}
                  \frac{ \sqrt{N_{L\uparrow,i}N_{R\downarrow,i}} -
                         \sqrt{N_{L\downarrow,i}N_{R\uparrow,i}}  }
                       { \sqrt{N_{L\uparrow,i}N_{R\downarrow,i}} + 
                         \sqrt{N_{L\downarrow,i}N_{R\uparrow,i}} },
\label{eqn}
\end{equation}
where $N_{L(R)\uparrow(\downarrow),i}$ is the number of events in 
the L (R) module when the beam polarization was up (down).
Equation (\ref{eqn}) cancels
spin dependent luminosity differences through second order.
Statistical errors were approximated by
$\Delta A_{N,i}=[P_b\sqrt{N_{L\uparrow,i}+N_{L\downarrow,i}+N_{R\uparrow,i}+N_{R\downarrow,i}}]^{-1}$,
valid for small asymmetries.
All measurements of $P_b$ for a store were averaged and applied 
to get $A_{N,i}$ for each bin.
The run-averaged $A_N\pm\Delta A_N$ values are 
shown in Fig.~\ref{fig:2}.

\begin{figure}
\includegraphics[height=1.45in]{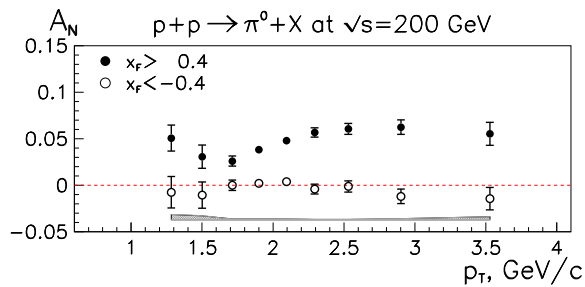}
\caption{Analyzing powers versus $\pi^0$ 
transverse momentum ($p_T$) for events with scaled $\pi^0$ longitudinal momentum
$|x_F| > 0.4$.  Errors are as described for Fig.~\ref{fig:2}.
\label{fig:3}}
\end{figure}

Systematic errors potentially arise from
several sources.  The bunch counter, used for the 
spin directions, identifies events in the abort gaps arising from
single-beam backgrounds.  They account for $<5\times10^{-4}$
of the observed yield.  Systematic effects from gain variations
with time are controlled by polarization reversals of the stored beam
bunches, as demonstrated by examples of spin-sorted $M_{\gamma\gamma}$
for L,R modules in the inset to Fig.~\ref{fig:2}.  Distributions
of the significance, $S_i=(A_{N,i}-A_N)/\Delta A_{N,i}$, are 
well described by zero mean value Gaussian distributions  
with $\sigma$ equal to unity, as expected if the uncertainties are
dominated by statistics, except near the trigger threshold where
larger $\sigma$ is observed.  Systematic errors are estimated 
from $\sigma\times\Delta A_N$ and differences in $A_N$
associated with $\pi^0$ identification, with the largest value chosen.
The upper limit on a correlated systematic error, common to all points,
arising from instrumental effects is $\delta A_N\approx 4\times10^{-4}$.

The same pair of modules concurrently measure $A_N$ values consistent with
zero for $x_F<0$ and $A_N$ that increases with $x_F$ for $x_F>0$,
depending on which beam spin is chosen.
%The $\langle\eta\rangle$=3.7 (3.3) data have $x_F>0$ for the Yellow (Blue) beam.
Null results at $x_F<0$ are natural since a possible gluon Sivers function is
probed where the unpolarized gluon distribution is large.  For $x_F>0$,
a calculation \cite{dalesio1,dalesio2} using quark Sivers
functions fit \cite{anselmino2} to SIDIS data \cite{hermes} best describes our
results at $\langle\eta\rangle=$3.3. Twist-3 calculations \cite{KQVY} that fit
$p_{\uparrow}+p\rightarrow\pi+X$ data at $\sqrt{s}=20$ GeV \cite{E704} and
preliminary RHIC results from the 2003 and 2005 runs at $\sqrt{s}$=200
GeV \cite{bland,brahms} best describe the data at 
$\langle\eta\rangle$=3.7.  Both calculations are in fair agreement with the
variation of $A_N$ with $x_F$.  Neither calculation describes data at both $\langle\eta\rangle$.

\begin{figure}
\includegraphics[height=2.72in]{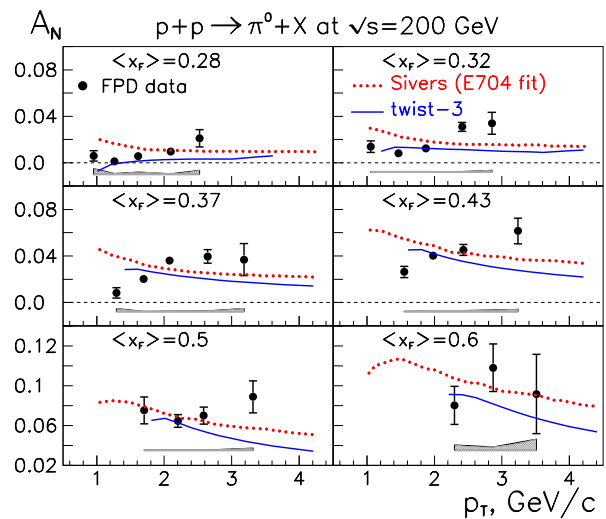}
\caption{Analyzing powers versus $\pi^0$
transverse momentum ($p_T$) in fixed $x_F$ bins (see Fig.~\ref{fig:1}).
Errors are as described for Fig.~\ref{fig:2}.  The
calculations are described in the text.
\label{fig:4}}
\end{figure}

Events from modules at different $\langle\eta\rangle$ that overlap
in the $x_F$--$p_T$ plane (Fig.~\ref{fig:1}) provide
consistent results.  Hence, it is possible to
further bin the results not only by $x_F$ but also by $p_T$.  For this
analysis, $p_T$ is determined from the measured energy, the
fitted position of the $\pi^0$ within an FPD module, and the measured 
position of the module relative to the beam pipe and to the
collision vertex.  The $z$ component of the event vertex uses a coarse time
difference between the east and west BBC, and is
determined to $\sim$20 cm resulting in $\Delta p_T / p_T$=0.04, where $\Delta p_T$ is the
uncertainty in $p_T$.  One method of determining
the $p_T$ dependence (Fig.~\ref{fig:3}) was to select events with $|x_F|>0.4$. 
$A_N$ is consistent with
zero for $x_F<-0.4$.  For $x_F>0.4$, there is a
hint of an initial decrease of $A_N$ with $p_T$, although the
statistical errors are large, since $\langle\eta\rangle$=4.0 data were only
obtained in the 2003 and 2005 runs with limited integrated luminosity
and polarization.  For $p_T>1.7$ GeV/c, $A_N$ tends to increase with
$p_T$ for $x_F>0.4$.  This is contrary to the theoretical expectation
that $A_N$ decreases with $p_T$.  

The results in Fig.~\ref{fig:3} may still reflect small correlations
between $x_F$ and $p_T$ for each point, rather than the dependence of
$A_N$ on $p_T$ at fixed $x_F$.  To eliminate this correlation, event
selection from Fig.~\ref{fig:1} was made in bins of $x_F$, followed by
bins in $p_T$.  The resulting variation of $A_N$ with $p_T$ is shown
in Fig.~\ref{fig:4}, compared to calculations \cite{dalesio1} using a Sivers function
fit to $p_{\uparrow}+p\rightarrow\pi + X$ data \cite{E704} and twist-3
calculations \cite{KQVY}.  For each point, the variation of $\langle x_F\rangle$ is
smaller than 0.01.  There is a clear tendency for $A_N$ to increase
with $p_T$, and no significant evidence over the measured range for
$A_N$ to decrease with increasing $p_T$, as expected by the
calculations.  This discrepancy may arise from unexpected TMD
fragmentation contributions, $x_F,p_T$ dependence of the requisite
color-charge interactions, evolution of the Sivers functions, or
from process dependence not accounted for by the theory.

In summary, we have measured the $x_F$ and $p_T$ dependence of
the analyzing power for forward $\pi^0$ production in $p_\uparrow+p$
collisions at $\sqrt{s}$=200 GeV in kinematics ($0.3<x_F<0.6$
and $1.2<p_T<4.0$ GeV/c) that straddle the region where cross sections are found in
agreement with pQCD calculations.  The $x_F$ dependence of the $\pi^0$
$A_N$ is in fair agreement with both a collinear twist-3
calculation and a calculation assuming factorization that attributes the
spin effects to spin-correlated intrinsic transverse momentum of the
quarks within the proton.  Recent theoretical work interrelates these
descriptions.  Both calculations expect the spin effects to
monotonically decrease with increasing $p_T$ for $p_T>$1.2 GeV/c.
Measurements of the $p_T$ dependence at fixed $x_F$ of $A_N$ 
are not consistent with these expectations.  This may reflect the
presence of additional mechanisms for these spin effects.  Future
measurements capable of disentangling TMD fragmentation and
distribution function contributions
to $\pi^0$ spin effects, and measurements of $A_N$ for real
and virtual photon production sensitive to only Sivers contributions,
are required to definitively establish if partonic orbital motion is
the correct explanation of these effects.

We thank the RHIC Operations Group and RCF at BNL, and the 
NERSC Center at LBNL and the resources provided by the 
Open Science Grid consortium for their support. This work was supported 
in part by the Offices of NP and HEP within the U.S. DOE Office 
of Science; the U.S. NSF; a sponsored research grant from Renaissance Technologies
Corporation; the BMBF of Germany; CNRS/IN2P3, RA, RPL, and 
EMN of France; EPSRC of the United Kingdom; FAPESP of Brazil; 
the Russian Ministry of Sci. and Tech.; the Ministry of 
Education and the NNSFC of China; IRP and GA of the Czech Republic,
FOM of the Netherlands, DAE, DST, and CSIR of the Government
of India; Swiss NSF; the Polish State Committee for Scientific 
Research; Slovak Research and Development Agency, and the 
Korea Sci. \& Eng. Foundation.

\end{document}